# Applied Theory of Mind and Large Language Models – how good is ChatGPT at solving social vignettes?

Anna Katharina Holl-Etten[1,+,*], Nina Schnaderbeck [1,+], Elizaveta Kosareva[1], Leonhard Aron Prattke[1], Ralph Krüger[2], Lisa Marie Warner[1], & Nora C. Vetter[1,*]

**Affiliations:**

[1] Faculty of Natural Sciences, Department of Psychology, MSB Medical School Berlin, Berlin, Germany

[2] Faculty of Information Science and Communication Studies/ Institute of Translation and Multilingual Communication, TH Köln – University of Applied Sciences, Cologne, Germany

[+]**Shared first authorship**

**\*Corresponding authors:**

Anna Katharina Holl-Etten and Nora C. Vetter

**Abstract**

The rapid development of language-based artificial intelligence (AI) offers new possibilities for psychotherapy and assistive systems, particularly benefitting autistic individuals who often respond well to technology. Parents of autistic persons emphasize the importance of appropriate and context-specific communication behavior. This study investigated whether GPT-3.5 Turbo and GPT-4, as language-based AI applications, are fundamentally capable of replicating this type of adequate communication behavior in the form of applied Theory of Mind (ToM). GPT-3.5 Turbo and GPT-4 were evaluated on three established higher-order ToM tasks: the Faux Pas Test, the Social Stories Questionnaire, and the Story Comprehension Test in English and German. Two independent raters scored response accuracy based on standardized manuals. In addition, responses were rated for epistemic markers as indicators of uncertainty. GPT's results were compared to human neurotypical and neurodivergent samples from previous own and others' research. GPT-4 achieved near human accuracy on the Faux Pas Test and outperformed GPT-3.5 Turbo and individuals with autistic traits. On the Social Stories Questionnaire, GPT-4 scored comparable to neurotypical adults, while GPT-3.5 Turbo remained well below. In the Story Comprehension Test, GPT-4 reached scores that exceeded neurotypical adult and adolescent benchmarks. However, GPT-4 used epistemic markers in up to 42% of responses. GPT-4 shows encouraging performance in complex higher-order ToM tasks and may offer future potential as an assistive tool for individuals with (and without) social communication difficulties. Its ability to interpret complex social situations is promising; however, the frequent use of uncertainty markers highlights the need for further study for assistive use and possibly further refinement to ensure consistent and reliable support in real-world use.

## Introduction

The increasing popularity and fast development of language-based artificial intelligence (AI) offers many potential applications in the field of clinical psychology (1). In particular, intervention technology that is assisted by AI for autistic persons shows great promise, given the frequent interest in technology among this group (2). Autism spectrum condition is a pervasive developmental disability characterized by persistent difficulties in social interaction and communication (3), for instance, taking ironic or sarcastic statements literally, problems with understanding faux pas situations, misunderstandings and other communication problems like understanding other persons' views. These and other difficulties mostly hinder connecting with others and lead to intense isolation including harassment in social contexts as they prevent autistic individuals from initiating and maintaining successful social interactions, for instance, with classmates, colleagues, etc. This can be attributed to autistic persons struggling with complex Theory of Mind (ToM), i.e., the ability to infer mental states, the beliefs, desires, intentions or emotions of others (e.g. 4,5). A prerequisite for complex social communication is a sophisticated higher-order ToM that comprises also other persons' use of faux-pas, irony or sarcasm (6). Altogether, autistic persons need lifelong support in adaptive behavior in different aspects of their daily life and social interactions (7). Relatedly, parents of autistic children emphasize that AI-based technologies may help to develop adequate and context-specific social communication skills (8). Consequently, it appears crucial to investigate the complex ToM abilities of large language models (LLMs) as a potential learning opportunity for them, and so that they could benefit from LLM-based assistance programs in everyday life which, e.g., help them interpret instances of irony, faux pas and sarcasm.

In contrast to neurotypical children, autistic children don't achieve certain developmental ToM milestones in the same way; sometimes they attain them at a later timepoint, only partially or not at all, respectively (9). Neurotypical children acquire a so-called understanding of *first-order false beliefs*, i.e. an understanding of true and false beliefs, between the ages of three and five (10). The understanding of the fact that others have different mental states and knowledge of the world to oneself can be tested using false belief tasks (11). Typically, these tasks are a version of the Unexpected Transfer paradigm (9), where, e.g., children are confronted with stories about two characters that have different information about the content of containers. For instance, character A moves an object from one container to another, but character B does not see this transfer. Consequently, one must infer that character B has different knowledge about the content than character A. The test question usually is: Is the ball/item in container A or B?

Based on first-order false belief tasks are so called *second-or higher-order false-belief tasks* (e.g., "she thinks that he knows that x knows where the object is") that encompass the understanding that one can have beliefs about another person's beliefs (12). Understanding of higher-order or so-called second-order false-belief develops between the ages 6 and 8 in neurotypical children and is crucial for complex social interactions, e.g. including jokes or deception.

Concerning other *complex, higher-order ToM abilities*, a number of such tasks have been developed, which involve a more mature understanding and use of complex language relating to intentions, emotions, and beliefs. One type of tasks are complex social vignette tasks; two examples are the "Strange Stories" (13), and the "Faux Pas" test (6) (14). They encompass short narratives that capture brief social encounters. For a complex understanding and high scores on these tasks, participants need to take the perspective of different protagonists to answer questions as to why they might think, feel, or behave as described in the narratives of the tests. The participants are typically asked in several steps, e.g. to indicate first whether an utterance includes an ironic statement or a social faux pas, whether the protagonist said something awkward, whether they could have known the related social facts about the other person and so forth. Therefore, not only complex social reasoning but also complex verbal abilities are required in several steps. These tasks have been developed to test the atypical development of higher-order ToM ability of autistic individuals who often struggle with these tasks (15). Regarding neurotypical development, these more advanced aspects of social reasoning typically develop in early adolescence (12,16). However, the question is whether LLMs could emulate these higher-order ToM tasks at a level comparable to that of neurotypical humans to support, e.g., autistic individuals who struggle with understanding these higher-order ToM abilities.

Correspondingly, research has recently begun to evaluate the ability of LLMs to solve different kinds of ToM tasks. First-order false belief tasks – typically solved by three- to four-year-old humans – are not solved well by GPT-3.5 (17) while GPT-4 (18) shows relatively strong performance with accuracies between approximately 70% and 97% (19,20). However, their reasoning remains fragile: minor variations in task structure or increased complexity can significantly reduce performance, indicating a reliance on pattern recognition rather than genuine belief attribution (21). In higher-order false-belief tasks, accuracies drop notably with GPT-4 achieving around 59-63% and GPT-3.5 performing below 50%, failing to correctly process recursive or nested belief structures and showing context-driven biases (22,23).

Most studies have focused on assessing LLMs' ToM abilities using first- and second-order false-belief tasks. These tasks are usually answered by a A or B / yes/no format ("is the item in box a or b; does person x know that person y knows this?"), making it easier to evaluate (correct/incorrect) the overall performance by LLMs in a large amount of (repeated prompt) data. Higher-order ToM tasks, particularly complex social vignette tasks, on the other hand require a coding scheme and the human evaluation (by two raters) of several phrases given by the LLM to the test questions. Therefore, these tasks are much more complex to evaluate and only a few recent studies have extended an LLM ToM benchmark to more complex social vignette tests. These include for instance the Hinting Task (24), which requires correctly identifying both the intended meaning of a remark and the action it aims to elicit, the Strange Stories paradigm (13), which requires interpreting the meaning of a story and providing a mentalistic explanation, and the Faux Pas Test (25), which consists of 20 social vignettes, of which ten contain a situation involving a faux pas, where one character unintentionally makes an inappropriate or awkward remark – typically due to the lack of knowledge or memory about relevant information – without any negative intent. The version by Stone et al. (1998) was developed for adults with frontal lobe lesions, whereas a corresponding child version was later introduced by Baron-Cohen et al. (1999) to assess ToM in autistic children.

GPT-3.5 achieved relatively high accuracies on narrative-based complex social vignettes, like the Strange Stories Task but struggled with inferring real intentions behind indirect speech utterances, as demonstrated by its poor performance on the Hinting Task of only 40-45% (26). Notably, the performance improved markedly, to 60-80% when additional cues or targeted questions regarding the characters' intentions were provided; however, in the majority of cases GPT-3.5 provided several hypotheses as an answer, which often included the correct answer, and in some cases, answers were qualified by remarks expressing doubts or uncertainty. Further, additional prompts were still necessary to obtain more accurate responses in the performance of GPT-4 in the Hinting Task (26). On average, GPT-4 reached only a deficient level of around 67% comparable to individuals with schizophrenia on the Hinting Task, who reached around 69% in an earlier study (27), and remarkably worse than neurotypical adults (performance around 80%). This suggests persistent limitations in the processing of implicit social cues. A study by Strachan et al. (2024) reported contrasting results in the comparison of GPT-3.5's and GPT-4's performance with a sample of human participants, presumably neurotypical adults. In their assessment, GPT-4 performed strongly in the Hinting Task (~100%), exceeding human and GPT-3.5's accuracy (~90%) contrary to the results by Brunet-Gouet et al. (2024). On the children's version of the Faux Pas Test (6), GPT-4 performed considerably worse with 60% in contrast to humans, who scored around 90% (28). GPT-3.5's performance was even weaker, scoring well below 50%, reflecting severe difficulties in correctly identifying unintentional social missteps. A notable error for faux pas was GPT-4's tendency to identify a statement as potentially offensive but failed to infer whether the speaker was aware of it. Notably, in Strachan et al.'s (2024) study, the main evaluation concentrated on one question from the Faux Pas test, the comprehension question. This question tests the understanding that the speaker did not remember the information that makes their statement inappropriate and is evaluated using the categories *correct* and *incorrect*, whereby the correct answer is always 'No, the speaker did not remember it'. However, additional analyses, where all four comprehension questions in accordance to Baron-Cohen et al.'s coding scheme (1999) were evaluated, resulted in the same picture (28). Importantly, their evaluation of the comprehension question excluded answers expressing uncertainty, which they coded as consideration of the correct answer but not committing to it. However, after alternative coding that included answers with uncertainty, the results of GPT improved only marginally (28). By contrast, on the Strange Stories Task, GPT-4 showed near-perfect performance outperforming human participants and GPT-3.5 by scoring around 80%. Despite GPT-4's overall strong performance, difficulties remain, particularly in detecting irony and sarcasm compared to humans.

Another study by Shapira et al., (2023) employed a modified version of the Faux Pas Test, also based on the child version (6) and comprising of a different set of vignettes developed by the authors. A comparable, though slightly higher, performance was observed, with GPT-3.5 achieving 73% and GPT-4 74% accuracy across the four questions following each vignette.

Furthermore, Attanasio et al.'s, (2024) research suggests that the new GPT-4 model achieves 100% accuracy comparable to neurotypical individuals on tasks assessing higher-order cognitive ToM, such as an Italian adapted version of the Strange Stories Task, called the Advanced ToM Task (30). However, GPT-3.5 achieved only 84%, and difficulties emerged in persuasion and double bluff scenarios, which required complex strategic reasoning. On tasks assessing affective ToM such as the Emotion Attribution Task (30), performance was even lower with GPT-3.5 and GPT-4 achieving only 53% and 65% accuracy, respectively, and performing similarly to autistic persons. In addition, the authors report that GPT tended to give verbose answers and needed more prompts to give precise answers (29). However, there was no additional evaluation of uncertain answers by GPT.

As reported in these previous studies, error analyses showed that LLMs often defaulted to cautious or vague interpretations that included uncertainty, or qualified their answers expressing doubts, revealing limitations in deeper social understanding (26,31). These kinds of statements can be called uncertainty markers, that is, in linguistic terms, "epistemic modalities" that are defined as "the speaker's assessment of probability and predictability" of a given state of affairs (32) 9). The use of epistemic markers by an LLM could indicate that the model expresses uncertainty and makes transparent the limits of its knowledge. When the answer of an LLM includes expressions such as *maybe*, *probably* or *possibly*, this may be showing that the model is not entirely sure of the accuracy of its response, which could be due to incomplete data or a question that goes beyond its training knowledge. This transparency about the basis of its answers, which, in LLMs, are based on probabilities rather than certainties, is particularly important in complex areas or domains with a high degree of uncertainty. The more uncertainty markers are included in an LLM's responses, the less useful it may be as a training or learning opportunity for individuals with social difficulties, such as autistic persons. Consequently, an evaluation of uncertainty markers seems vital for a complete usability assessment of LLM's in an assistive context.

**Aims of the current study**

Given the current inconsistencies and the limited number of studies investigating complex social vignettes, the present study aimed to systematically assess GPT's performance on three tasks taken from research on autism spectrum disorder, the Faux Pas Test (25), the Social Stories Questionnaire (33), and the Story Comprehension Test (34), designed to capture different aspects of complex social interactions. Furthermore, we evaluated a more complex version of the Faux Pas Test (25) than previously tested in LLM's such as GPT (28) containing not only the 4 test questions of the original manual (25) to further challenge the model's social reasoning abilities, but also two additional questions that are part of the German version used in the present study (35): an *empathic understanding* question ("How do you think X felt?") and a *false-belief* question assessing whether the respondent recognized that the faux pas was unintentional. Going beyond previous research, the test questions were not limited to answers that could simply be rated as false or correct. For the tasks standardized evaluation manuals are available and more intensive two-rater evaluation is mandatory in contrast to most previous studies that, e.g., evaluated false belief reasoning. We also evaluated whether the answers contained vague or cautious interpretations that can be specified as uncertainty markers. This is an important factor for potential applications as an assistive tool for autistic individuals, who benefit most from clear and unambiguous responses. In line with previous research, we tested both GPT-3.5 and GPT-4 to compare their performance.

In contrast to previous research, we tested both English and German to explore potential differences related to the language of input, given that GPT's training data is substantially larger in English (36). This approach also addresses concerns about contamination of our results through training, as all three tasks were originally developed in English and were probably administered to GPT in this language before. Moreover, testing in the native language of users reflects real-world usage scenarios.

Altogether, the aim of our study is to determine whether widely used LLM's, here GPT-3.5 and GPT-4, are able to provide this kind of adequate and context-specific social communication assistance by testing applied higher-order ToM, as required by individuals with problems in social communication such as autistic persons (8)

**Method**

**Procedure and Data Collection**

The performance of GPT version 3.5 and version 4 was explored by posing the short social stories and follow-up questions of three well-established ToM tests: the Faux Pas Test (25), the Social Stories Questionnaire (SSQ; 33) and the Story Comprehension Test (34). All tests are validated in various clinical and non-clinical populations (37,e.g. 38). All tests were applied in English and German. Since no validated German version of the SSQ existed, two master's students translated it from English. Subsequently a native German speaker retranslated it and the translations were compared and harmonized. For the Faux Pas Test we used the validated German version by Ströbele (2004) and for the Story Comprehension Test, our validated German translation by Vetter et al. (2013). All the GPT outputs were rated by two independent raters. The ratings were strictly orientated according to the original manuals. Divergent ratings of correctness and uncertainty markers were resolved through discussion based on original scoring criteria (interrater reliability κ = 0.91–0.97). From these, an overall score was computed. Means and standard deviations were then calculated for the tests, languages, and versions. Also, we compared GPT's results from previous studies with neurotypical participants and participants with impaired ToM abilities, such as autistic individuals or participants with autism traits. GPT would have to perform at the level of neurotypical participants to potentially provide assistance in the form of digital mental health applications for autistic persons or other individuals with social communication difficulties (8).

We collected the data in November 2023 using the OpenAI API (Application Programming Interface). To access the API, a subscription to OpenAI was set up and secured with a personal API key that serves as an authentication tool. A Python script (S1 Appendix) was used that performed the following tasks: read the questions from the spreadsheet, create API requests from the questions, submit them to the OpenAI API server, and extract the responses. Each ToM test was prompted in both English and German. To obtain reliable results, each ToM test was queried ten times per condition. A built-in error handling mechanism ensured that if no valid response was received within 60 seconds, the query would automatically be resent.

When formulating the prompts (S2 Appendix), prompt engineering was implemented to minimize hallucinations and inaccurate or rambling answers, because in pilot testing GPT usually answered very inaccurate and provided several potential possibilities.

**Measures**

***Faux Pas Recognition Test***

The Faux Pas Recognition Test was originally developed to assess ToM abilities in autistic persons (25). The test consists of 20 social vignettes: ten describe a social situation involving a faux pas, while the remaining ten serve as control stories that describe social interactions but do not contain a faux pas. The stories which contain a faux pas are presented at specific positions in the test (2;4;7;11;12;13;14;15;16;18). In the faux pas vignettes, one character unintentionally makes an inappropriate or awkward remark – typically due to the lack of knowledge or memory about relevant information – without any negative intent. In the validated German version (35) each story is followed by eight questions. The first six questions assess social reasoning and emotional understanding (i.e., ToM), while the last two are control questions addressing factual details. The structure is illustrated by the following example vignette:

> "Helen's husband was throwing a surprise party for her birthday. He invited Sarah, a friend of Helen's, and said, 'Don't tell anyone, especially Helen.' The day before the party, Helen was over at Sarah's and Sarah spilled some coffee on a new dress that was hanging over her chair. 'Oh!' said Sarah, 'I was going to wear this to your party!' 'What party?' said Helen. 'Come on,' said Sarah, 'Let's go see if we can get the stain out.'"

The following questions are asked:

1. Did anyone say something they shouldn't have said or something awkward? (Faux pas detection)

2. Who said something they shouldn't have said or something awkward? (Identification question)

3. Why shouldn't he/she have said it or why was it awkward? (Understanding of inappropriateness)

4. Why do you think he/she said it? (Intention question)

5. Did Sarah remember that the party was a surprise party? (Belief question)

6. How do you think Helen felt? (Empathy question)

7. What got spilled on the dress? (Control question)

8. In the story, who was the surprise party for? (Control question)

For each faux pas vignette, if the first question (faux pas detection) was answered correctly, one point was given for each correct answer of the six questions, resulting in a maximum of 6 points per vignette and a total of 60 points across all ten faux pas stories. In addition to correctness, raters also evaluated whether an uncertainty marker was mentioned in response to each question. Responses were scored one point if a wording such as "maybe" was used, or zero if no uncertainty marker was used.

### *Social Stories Questionnaire (SSQ)*

The Social Stories Questionnaire (SSQ) (33) consists of 10 social vignettes, which describe social interactions. During the interaction one of the characters makes potentially offensive or hurtful remarks. Each short story is divided into three sections (A, B&C). The resulting 30 items are designed to measure the detection of upsetting comments, depending on the severity of the upsetting comment. In every section there is a short dialogue with equivocal statements. While 10 sections are very subtle and hard to notice, another 10 sections are more obvious. Lastly, there are ten sections that don't contain any upsetting comments. At first, it should be decided whether the text contains a potentially upsetting sentence at all. In case that there is one, it must be specified which one it is, and it must be identified whether the remark is in fact upsetting to the second character. An example of section A of story seven (blatant faux pas) is:

> "Linda was feeling nervous. Tonight, would be the first time that her good friends Alex and Faye were going to meet her new boyfriend, James. In part this nervousness was because Alex and Faye had been very keen on Roger, her last boyfriend. However, Linda had also received a phone call from her old best friend Kate, the night before. Kate and Linda had been best friends at college. However, after Kate had been offered the job in Moscow they only kept in touch by post. This was the way it had been for the last five years but that was going to change tonight. Kate was back in London for the weekend and coming to dinner. Just after 7 pm the doorbell rang. Linda opened the door to see her old friend Kate smiling back at her.
>
> o "Hello stranger," said Linda giving Kate a big hug, "come on in".
>
> o "My goodness" said Kate, "there are a few more lines on that face than I remember."
>
> o "Well, it has been five years," replied Linda. "So back in the old city huh.
>
> o "What's brought you back to London?"
>
> Question: Was anything said in the previous section that could have upset someone?
>
> If yes, indicate what line it occurred in by filling in the circle. If no, please proceed to the next.

The ranking system is based on the correct solutions following Lawson et al. (2004). For each correctly identified offensive comment, one point is awarded, so that a maximum of 20 points can be achieved. A sum score was calculated, by summing up the number of correct line identifications for the ten subtle and the ten blatant story sections. The individual sum scores for subtle and blatant faux pas are presented in S3 Appendix.

### *The Story Comprehension Test (SCT)*

The Story Comprehension Test (SCT) was developed by Channon and Crawford (2000) to research the ToM capabilities of people with posterior and anterior brain lesions. Our lab is experienced in assessing this task since we already used it in a study comparing adolescents with adults (16). For this work the slightly modified and translated version of the test by Vetter et al. (2013) was used.

We followed up on this research using seven out of the original 12 short vignettes, because our prior pilot testing revealed that certain vignettes were culturally inappropriate for a German study sample, for example the vignette involving queuing at bus stops in the UK was excluded (16). The short stories describe an interaction with a dialog of two people. After each story the participant is asked to explain why a character said or did something. To correctly answer this question, advanced mentalizing abilities are required in order to recognize the complex social intentions in the vignettes, such as pretense, white lie, irony, threat, and dare. For each vignette a specific set of grading criteria with examples of possible responses was created. The responses are classified as correct (2 points) if words or actions of the story character were interpreted accurately and completely, as partly correct (1 point) if the interpretation was correct but not complete, or as incorrect (0 points) if the mental state was notrecognized. This is an example item:

> "Michael goes to his friend Amanda's wedding reception. Amanda and Michael have been good friends for a long time. Michael is worried because he doesn't think she and her husband are a good fit for each other and believes the marriage won't last. After the wedding, Michael goes to Amanda and says: 'Such a wonderful day, I'm sure you'll be very happy together.' Why did Michael say that?"

For each vignette, there is one test question, consequently, a maximum amount of 14 points is possible for the seven questions.

**Uncertainty Markers**

In addition to the original test criteria raters also evaluated whether an uncertainty marker was mentioned in the responses of the Faux Pas Test and the Story Comprehension Test. Uncertainty markers are scored dichotomously. If a response contained one, a point was noted; if no uncertainty marker was used, zero points were given. The proportion of responses that include uncertainty markers is displayed separately from the overall results. In contrast to the mentioned tests, the occurrence of uncertainty markers in the Social Stories Questionnaire cannot be measured as a dependent variable, since the task itself already encompasses such modalities. Their presence is consistently observed in the responses and should therefore not be considered inherently significant.

**Statistical Analysis**

To assess the accuracy of the two GPT versions, the responses to each test were scored in accordance with the predefined evaluation criteria. Accuracy was reported as a percentage for each test. Considering missing data, Story 16 of the Faux Pas Test was excluded from the analysis due to systematic missing responses caused by an unclear error in the results provided by GPT.

Group comparisons were conducted depending on assumption checks (Shapiro–Wilk test for normality, Levene's test for homogeneity of variances). If assumptions were met, two-way ANOVAs were applied; in cases of unequal variances, Welch's ANOVA was used. When assumptions for parametric testing were violated, pairwise comparisons between GPT versions were carried out using Wilcoxon signed-rank tests. Effect sizes were reported as partial eta squared ($\eta^2$) for ANOVA's and correlation coefficients ($r$) for Wilcoxon tests.

## Results

The following section presents the results of the Theory of Mind test battery, which included the Faux Pas Test, the Social Stories Questionnaire, and the Story Comprehension Test. Figure 1 compares the performance of GPT-3.5 and GPT-4 across all tests in the battery. In addition, the respective performances of GPT versions are compared with those of neurotypical individuals. For the Faux Pas Test and the Story Comprehension Test, the analysis of uncertainty markers is also reported.

**Figure 1**

*Performance of GPT on the battery of Theory of Mind Tests*

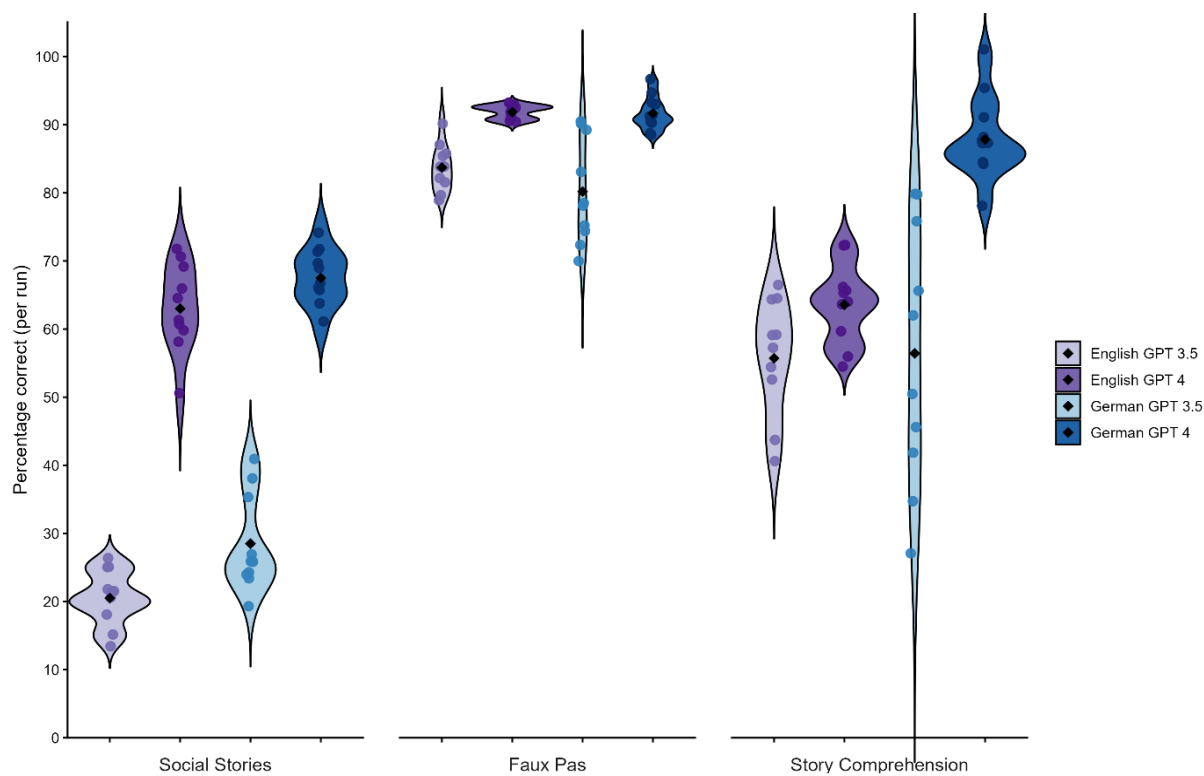

*Note.* GPT-3.5 English (light purple), GPT-4 English (dark pruple), GPT-3.5 German (light blue) and GPT-4 German (dark blue). The plot shows the distribution of percentage-correct scores for each task, based on 10 runs per condition. Coloured dots represent the percentage-correct score for a single run (i.e., averaged across all stories within that run). Black diamonds indicate the mean percentage-correct score across all runs for each condition.

## Faux Pas Test

The lowest average point score across runs was observed for the German prompts to GPT in version 3.5 ($M$= 43.30, $SD$ = 4.16), with a higher score for the English prompts in the same version ($M$= 45.20, $SD$ = 1.87). The highest average point score was achieved in English prompts to GPT-4 ($M$ = 49.60, $SD$ = 0.52), with a slightly lower score for the German prompts in the same version ($M$ = 49.50, $SD$ = 1.27). Table 1 displays the total points achieved for each story, aggregated across all 10 runs.

A Wilcoxon signed-rank test showed a significant difference between the scores for the versions GPT-3.5 and GPT-4, for both English prompts ($z = -2.67$, $p = .009$, $r = .89$) and German prompts ($z = -2.52$, $p = .014$, $r = .89$). The effect sizes indicate strong differences between the two model versions. The comparison of GPT's performance on the faux pas test with neurotypical and adults with subclinical autism symptoms is presented in Figure 2.

**Figure 2**

*Comparison to Human Test Performance in Faux Pas Test (39)*

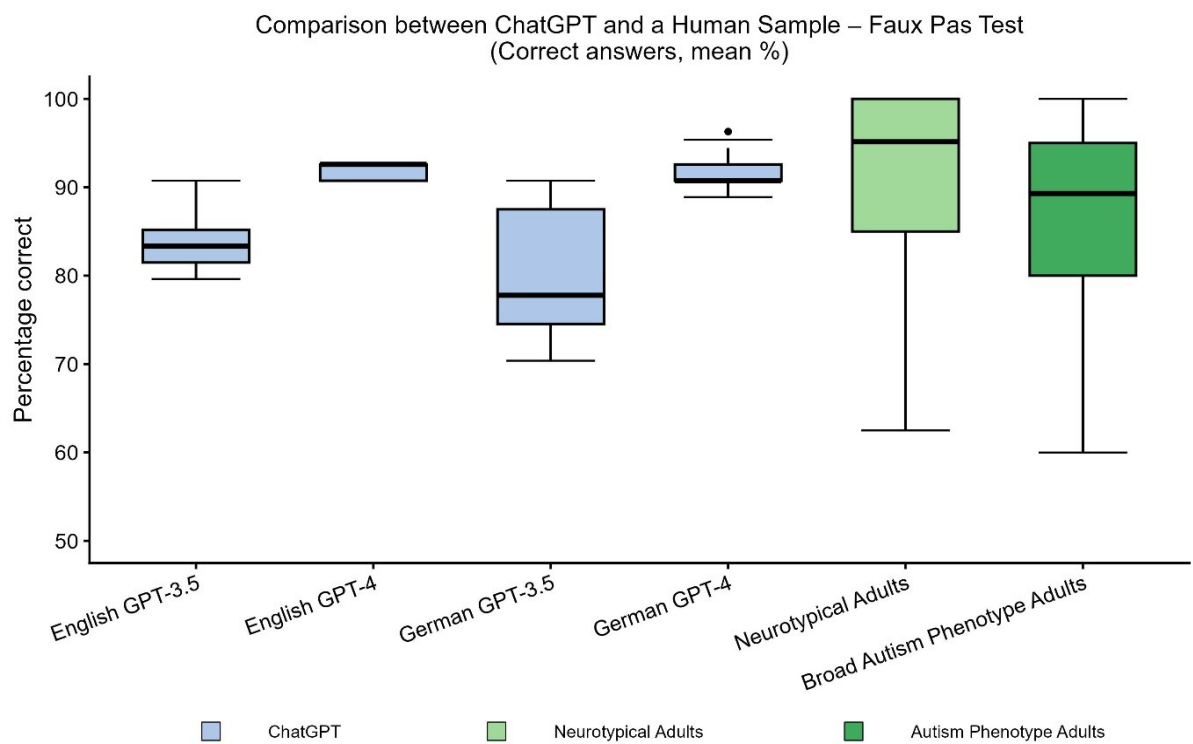

*Note.* Means and errors (%) for neurotypical (*n*= 59) and adults with subclinical autism symptoms (*n*= 59) as found by Atherton and Cross (2019). GPT-4 was able to detect the presented faux pas equally well in both languages (92% German; 92% English)– at a level comparable to neurotypical adults (95%) – while GPT version 3.5 showed notable limitations (80% German; 84% English). Its performance pattern resembled that of individuals on the Broad Autism Phenotype (89%) or even below.

**Table 1**

*Points and Percentage correct of the Faux Pas Test per story in both languages and both GPT versions*

| Language/ Version | Score | Story 2 (Birthday) | | Story 4 (Apartment) | | Story 7 (Mistake) | | Story 11 (Joke) | | Story 12 (Slander) | | Story 13 (Cake) | | Story 14 (Present) | | Story 15 (Competition) | | Story 18 (Lawyer) | |
|---|---|---|---|---|---|---|---|---|---|---|---|---|---|---|---|---|---|---|---|
| | | P | % | P | % | P | % | P | % | P | % | P | % | P | % | P | % | P | % |
| Eng.3.5 | P | 56 | 93 | 49 | 82 | 53 | 88 | 47 | 78 | 60 | 100 | 58 | 97 | 45 | 75 | 49 | 82 | 35 | 58 |
| Eng.4 | P | 57 | 95 | 50 | 83 | 60 | 100 | 53 | 88 | 59 | 98 | 55 | 91 | 60 | 100 | 50 | 83 | 52 | 87 |
| Ger.3.5 | P | 60 | 100 | 50 | 83 | 47 | 78 | 43 | 72 | 58 | 97 | 49 | 82 | 40 | 67 | 50 | 83 | 36 | 60 |
| Ger.4 | P | 59 | 98 | 50 | 83 | 60 | 100 | 54 | 90 | 56 | 93 | 53 | 88 | 60 | 100 | 50 | 83 | 53 | 88 |
| Eng.3.5 | EpM | 15 | 19 | 20 | 25 | 17 | 21 | 18 | 23 | 20 | 25 | 20 | 25 | 17 | 21 | 19 | 24 | 15 | 19 |
| Eng.4 | EpM | 27 | 45 | 27 | 45 | 20 | 25 | 19 | 24 | 20 | 25 | 20 | 25 | 21 | 35 | 20 | 25 | 20 | 25 |
| Ger.3.5 | EpM | 24 | 40 | 23 | 38 | 17 | 21 | 21 | 35 | 21 | 35 | 20 | 25 | 21 | 35 | 20 | 25 | 20 | 25 |
| Ger.4 | EpM | 30 | 38 | 26 | 33 | 25 | 31 | 21 | 35 | 29 | 48 | 22 | 28 | 22 | 28 | 27 | 34 | 23 | 38 |

*Note.* Displayed as "Number of Points; "Percentage of total Points". Displayed in separate rows are; P = Points, EpM = uncertainty marker. The displayed stories are the stories containing faux pas. Story 16 was excluded due to systematic missings. Achieved scores in the Faux Pas Test, summarized separately for each story. Values represent the total number of correctly identified faux pas across 10 runs per condition (maximum = 60 points per story). Percentages indicate the proportion of correct responses relative to the maximum possible score per story.

***Results of Uncertainty Markers in Faux Pas Test***

The occurrences of uncertainty markers showed a different pattern than the test scores, with GPT-4 displaying a higher frequency of uncertainty markers, regardless of version and language. The fewest average mentions of uncertainty markers per run (Figure 3) were observed for the English prompts to GPT-3.5 ($M = 16.10$, $SD = 0.74$), with a higher occurrence in response to the German prompts in the same version ($M = 18.70$, $SD = 2.11$). The highest occurrence was observed in response to the German prompts in GPT-4 ($M = 22.5$, $SD = 1.35$), with a lower number of occurrences for the English prompts in the same version ($M = 19.40$, $SD = 0.84$). A Wilcoxon signed-rank test revealed a significant difference in the occurrence of uncertainty markers between GPT-3.5 and GPT-4, for both English prompts ($z = -2.80$, $p = .006$, $r = .89$) and German prompts ($z = -2.80$, $p = .006$, $r = .89$).

**Figure 3**

*Occurrence of Uncertainty Markers in the Faux Pas Test*

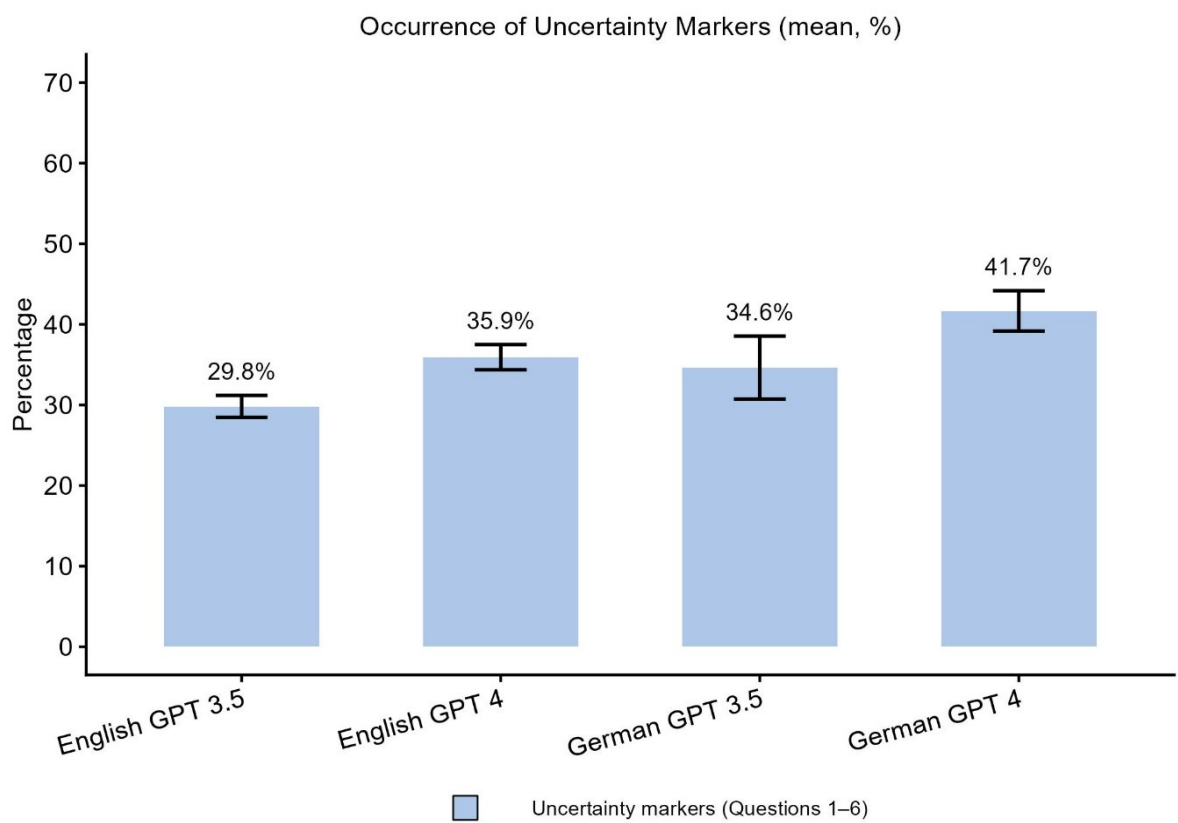

*Note.* Count of answers per trial with at least one uncertainty marker, across versions and languages in the Faux Pas Test.

**Social Stories Questionnaire**

The lowest average point score across runs was observed for the English prompts to GPT in version 3.5 ($M = 4.10$, $SD = 0.74$), with a higher score for the German prompts in the same version ($M = 5.70$, $SD = 1.42$). The highest average point score was achieved in German prompts to GPT in Version 4 ($M = 13.50$, $SD = 0.85$), with a slightly lower score for the English prompts in the same version ($M = 12.60$, $SD = 1.26$).

A two-way ANOVA was conducted to examine the effect of language and version on the accuracy of GPT in the Social Stories Questionnaire. The main effect of language was statistically significant, $F(1,36)= 10.50$, $p= .003$, partial $\eta^2= .26$. This indicates that the German prompts led to significantly higher scores compared to the English prompts. The main effect of version was also statistically significant, $F(1,36)= 249.46$, $p < .001$, partial $\eta^2= .94$, suggesting that the version factor had a strong effect on performance, with GPT-4 outperforming GPT-3.5. However, the interaction effect between language and version was statistically not significant, $F(1,36)= 1.00$, $p= .323$, partial $\eta^2= .03$. Figure 4 presents a comparison between GPT's performance and that of neurotypical adults and males with Asperger syndrome on the Social Stories Questionnaire.

**Figure 4**

*Comparison GPT to Human Test Performance in the Social Stories Questionnaire (Lawson et al.,2004)*

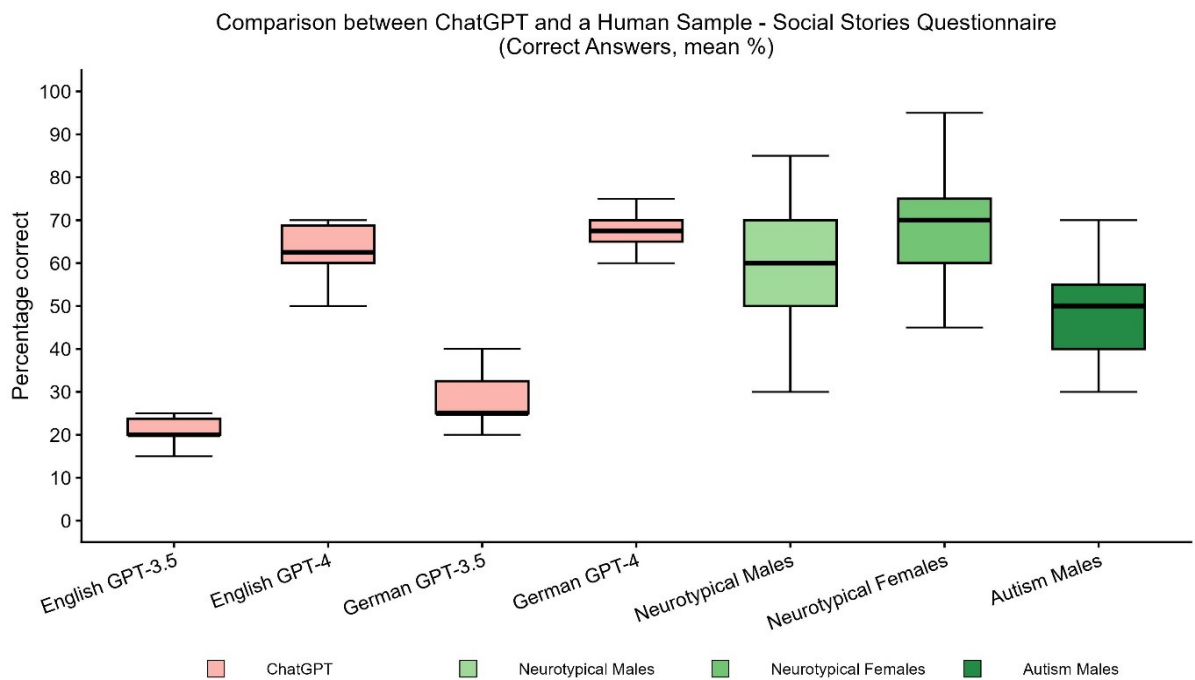

*Note.* Means and errors (%) for neurotypical males (*n*= 44), neurotypical females (*n*= 45) and males with Asperger syndrome (*n*= 18) as found by Lawson et al., (2004). Regardless of the language in which the items were prompted, GPT-4 (67% German; 63% English) exhibited a test score comparable to neurotypical adults (70% Females, 60% Males) from Lawson et al., 2004. In contrast, the performance of GPT-3.5 was below the level reported for neurotypical adults and male subjects with an Asperger's diagnosis (50%) in both languages (28% German; 20% English).

**Story Comprehension Test**

The lowest average point score across runs was registered for the English prompts to GPT-3.5 (*M* = 7.80, *SD* = 1.14), with a slightly higher score for the German prompts in the same version (*M* = 7.90, *SD* = 2.64). GPT-4 achieved higher average point scores in both languages, German (*M* = 12.3, *SD* = 0.82) and English (*M* = 8.9, *SD* = 0.74).

A Welch's ANOVA revealed a significant effect of group on the number of correct responses in the Story Comprehension test, $F(3, 19.2) = 44.60$, $p < .001$, $\eta^2 = 0.61$. Games–Howell post hoc comparisons indicated that the German GPT-4 outperformed GPT-3.5 both, the German (*p* = .002) and English prompts (*p* < .001). Moreover, GPT-4 achieved significantly higher scores in the German prompts than in the English prompts ($p < .001$). In contrast, no significant differences were found between German and English prompts in GPT-3.5 ($p = .999$), between German GPT-3.5 and English GPT-4 ($p = .668$), or between GPT-3.5 and GPT-4 in the English prompts ($p = .088$). Table 2 displays the achieved scores separately for each story, summed across all runs. Figure 5 compares the performance of GPT with that of a human sample from previous work of our research group (16).

**Table 2**

*Points and Percentage correct of the Story Comprehension Test per Story in both Languages and both GPT Versions*

| Language/Version | Score | Story 1 (Pretense) | | Story 2 (Threat) | | Story 3 (Irony) | | Story 4 (Dare) | | Story 5 (White Lie) | | Story 8 (Excuse) | | Story 10 (Elaborated Irony) | |
|---|---|---|---|---|---|---|---|---|---|---|---|---|---|---|---|
| | | P | % | P | % | P | % | P | % | P | % | P | % | P | % |
| Eng.3.5 | P | 10 | 50 | 15 | 75 | 20 | 100 | 7 | 35 | 13 | 65 | 10 | 50 | 3 | 15 |
| Eng.4 | P | 11 | 55 | 10 | 55 | 10 | 55 | 14 | 70 | 13 | 65 | 11 | 55 | 20 | 100 |
| Ger.3.5 | P | 12 | 60 | 10 | 50 | 6 | 30 | 15 | 75 | 16 | 80 | 17 | 85 | 3 | 15 |
| Ger.4 | P | 16 | 80 | 16 | 80 | 20 | 100 | 11 | 55 | 20 | 100 | 20 | 100 | 20 | 100 |
| Eng.3.5 | EpM | 1 | 10 | 0 | 0 | 6 | 60 | 0 | 0 | 3 | 30 | 0 | 0 | 0 | 0 |
| Eng.4 | EpM | 1 | 10 | 4 | 40 | 10 | 100 | 2 | 20 | 0 | 0 | 1 | 10 | 3 | 30 |
| Ger.3.5 | EpM | 6 | 60 | 4 | 40 | 3 | 30 | 6 | 60 | 3 | 30 | 8 | 80 | 5 | 50 |
| Ger.4 | EpM | 2 | 20 | 0 | 0 | 7 | 70 | 3 | 30 | 2 | 20 | 3 | 30 | 2 | 20 |

*Note.* Displayed as "Number of Points; "Percentage of total Points". Displayed in separate rows are; P = Points, EpM = uncertainty marker. Presents the achieved total scores per story in the Story Comprehension Test, aggregated across all 10 runs (maximum = 20 points per story). Percentages indicate the proportion of the maximum achievable score.

**Figure 5**

*Comparison to Human Test Performance found previously in our own research group (16) in the Story Comprehension Test*

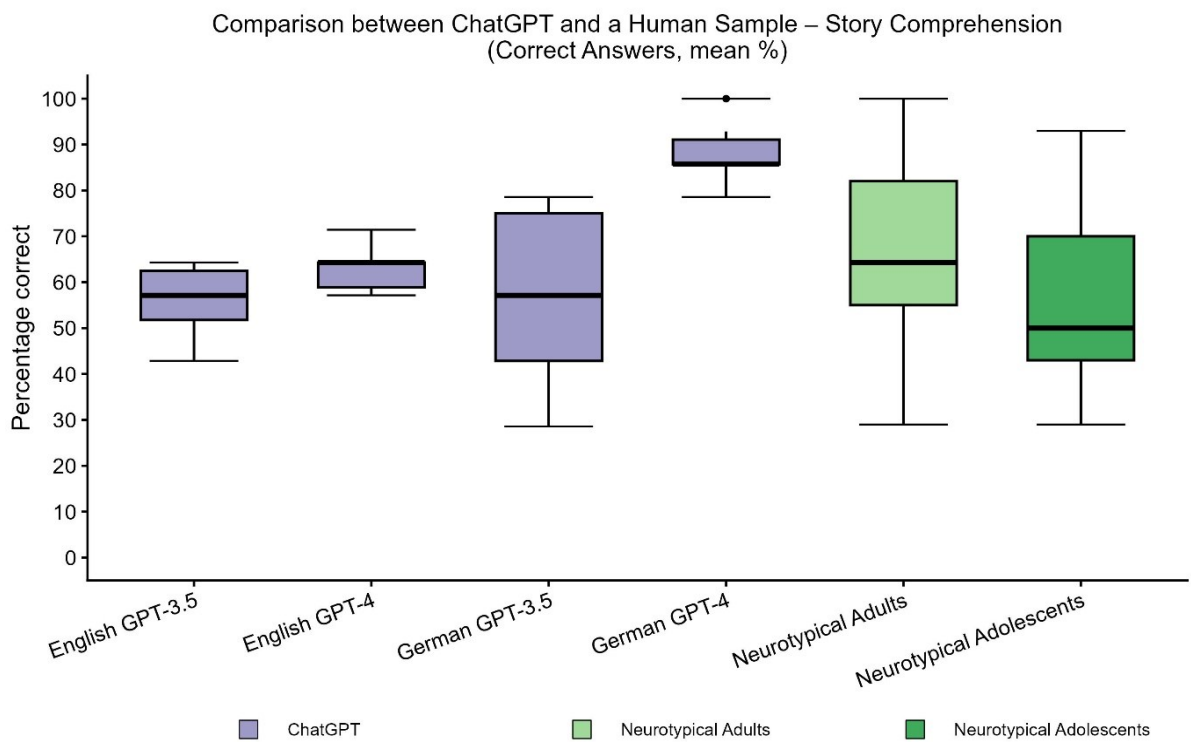

Comparison between ChatGPT and a Human Sample – Story Comprehension
(Correct Answers, mean %)

*Note.* Means and errors (%) for neurotypical adults (*n*= 60) and neurotypical adolescents (*n*= 60) as found by Vetter et al., (2013). GPT-4 in the German version achieved an accuracy (89%) that exceeded that of both neurotypical adults (64%) and adolescents (50%). In the English version, accuracy (64%) was lower, but still comparable to neurotypical adults and higher than that of neurotypical adolescents. By contrast, GPT-3 showed lower accuracy (56% German and English) across both languages and versions, however, its performance still reached a level similar to that of neurotypical adults.

### Uncertainty Markers in Story Comprehension Test

The occurrences of uncertainty markers showed another picture compared to the test scores. Similarly, the fewest average mentions of uncertainty markers per run were observed for the English prompts to GPT-3.5 ($M$ = 1.0, S$D$ = 0.82), however, followed by a slightly higher occurrence in responses to the German prompts in version GPT-4 ($M$ = 1.9, $SD$ = 1.29). Higher occurrences resulted in responses to the English prompts in GPT-4 ($M$ = 2.1, $SD$ = 0.57) and in responses to the German prompts in GPT-3.5 ($M$ = 3.5, $SD$ = 1.51).

A two-way ANOVA revealed a significant main effect of language, $F(1,36)$ = 25.40, $p$ < .001, $\eta^2$ = .23, indicating that the number of uncertainty markers differed between German and English prompts. There was also a significant main effect of version, $F(1,36)$ = 10.40, $p$ =

.003, η² = .01, suggesting differences between GPT-3.5 and GPT-4. Most importantly, the interaction between language and version was significant, $F(1,36) = 14.81$, $p < .001$, η² = .29, showing that the effect of version on uncertainty markers depended on the prompting language. Figure 6 shows the comparison between the occurrence of uncertainty markers made from GPT and a human sample from previous work of our research group (16).

**Figure 6**

*Comparison of the Occurrence of Uncertainty Markers between GPT and a Human Sample in the Story Comprehension Test (16)*

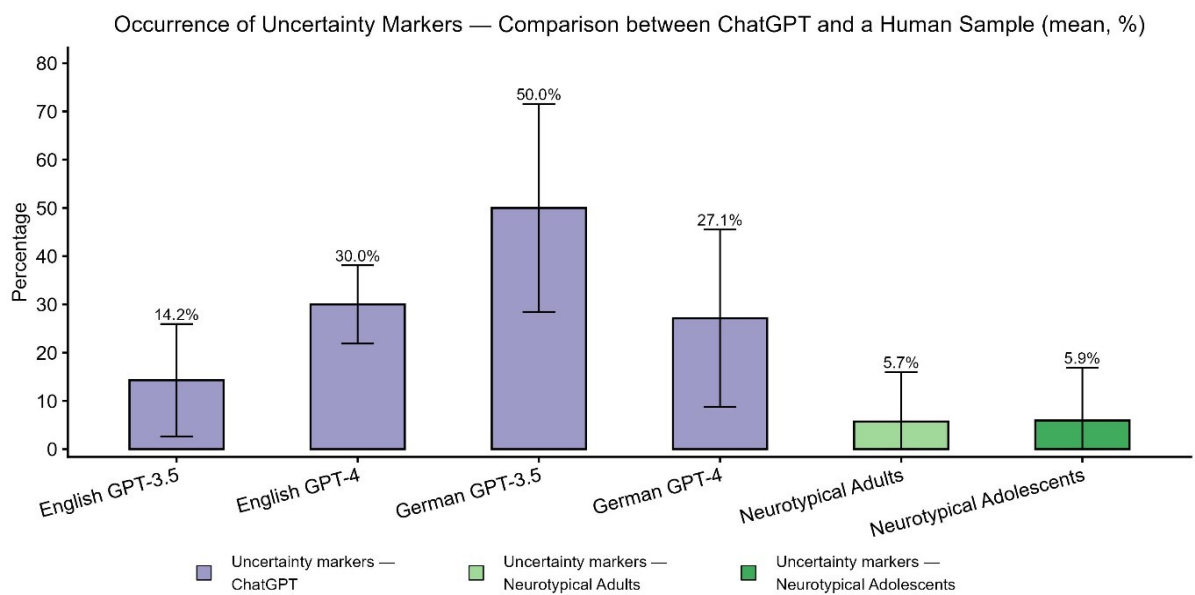

*Note.* GPT: Count of answers per trial with at least one uncertainty marker, across versions and languages in the Story Comprehension Test. Human Sample: Count of answers per trial with at least one uncertainty marker among $n= 60$ neurotypical adults and $n= 60$ neurotypical adolescents in the Story Comprehension Test (16).

## Discussion

In the few existing previous studies (26,28,29), LLMs' performance in complex higher-order ToM tasks was mixed. Therefore, one objective of this study was to systematically evaluate GPT's ability to solve higher-order ToM tasks, i.e. complex social vignettes, and thus to determine its potential as an assistive tool for individuals with social communication difficulties. Three established measures of complex social reasoning were employed: the Faux

Pas Recognition Test, the Social Stories Questionnaire, and the Story Comprehension Test in both English and German, these were assessed for accuracy and uncertainty markers.

**Faux Pas Test**

In the faux pas test, GPT-4 performed well, achieving nearly full scores across all iterations in both German and English. Overall, the performance of GPT-4 can be described as on par with neurotypical adults (95%; Atherton and Cross, 2019) in both languages (92% German; 92% English); albeit, the previous version GPT-3.5 showed notable limitations (80% German; 84% English). Its performance pattern was even below that of individuals on the Broad Autism Phenotype (89%), when compared with the findings of Atherton and Cross (2019). However, GPT-4 performed notably better than in the study by Strachan et al. (2024), where it reached only an accuracy of 60%, and also slightly higher than in the study by Shapira et al. (2023) with GPT-3.5 achieving 73% and GPT-4 74%; albeit, they applied a modified version of the Faux Pas Test comprising of a different set of vignettes.

Different reasons could be accountable for this discrepancy; first, prompt engineering could be one reason. As prompting issues were mentioned in previous studies (26; 28), we instructed GPT to decide on a single most likely answer, to answer as briefly and precisely as possible, as well as to 'put itself in the shoes of the characters' and adopt their perspective and intentions. In contrast, Strachan et al. (2024) merely informed GPT within the prompts that, at the end of each story, a person would say or do something, followed by a question about what had happened in the story. Secondly, in our study we used the German version of the Faux Pas task (35) which had slightly different questions. Specifically, we had six scoring questions and two control questions in contrast to four questions plus one control question in the English version (25) that was administered in the study by Strachan et al. (2024). We used the version of the Faux Pas Task (35) with six questions in both languages, German and English, to allow for a cross-linguistic comparison, thus, this could have made a difference. Thirdly, as was reported by Strachan et al. (2024), they scored all responses as incorrect, when they indicated some uncertainty, specifically regarding answers to the comprehension question, in contrast to our coding that registered uncertainty separately. However, in additional analyses they showed that the results remained largely similar when also considering answers that contained uncertain statements, but they 'hyperconservative approach' could still have made a difference (28). Finally, another relevant point regards the specific time-point and language version of GPT-4, as our study took place after Strachan et al.'s (2024), this could have resulted in a higher amount of training data for GPT in the meantime, specifically in the English version; albeit, our results in the German version were also considerably higher than theirs'.

Considering the evaluation of uncertainty markers, our results showed a different picture with the highest amount of 42% for German prompts in GPT-4 and the lowest amount of 30% for the English prompts in version 3.5. The results suggest the assumption that the usage of uncertainty markers increased in version 4, that might be a consequence of the policy of OpenAI to improve usability and transparency of GPT's certainty to relative answers (18). As indicated above, GPT might indicate here that it is not sure of the accuracy of its answer or that the question goes beyond its training. As this study is the first to systematically evaluate uncertainty markers of LLM's in ToM stories it is unclear whether other LLM's show similar response patterns in uncertainty markers also known as "linguistic epistemic modalities". However, the applicability of GPT in real-world scenarios is reduced notably by this large number of weakening statements in its answers. This is because neurodivergent individuals would need to decide for themselves whether certain statements offer a plausible explanation for ambiguous social situations. Consequently, the clear guidance that neurodivergent individuals require would need further refinement in at least one third of GPT's answers relating to Faux Pas, which emphasizes the need for further studies including neurodivergent and neurotypical persons who should evaluate the helpfulness in social situations (8).

Regarding our observation that the number of uncertainty markers is considerably higher in German versus English language prompts, a comparison to other studies is not yet possible due to the lack of existing evidence for German or LLM prompts in general in ToM tasks (40). One reason for the higher amount could be that GPT had more training with the administered tasks in English, as mentioned above, possibly because of previous studies that were mainly administered in English or as a consequence of their inclusion in model training (Chen et al., 2024). As GPT had the highest amount of general training in the English language (Lommel, 2024), thus, this could lead to more expressions that contain uncertainty in German language responses, because in more recent versions GPT is programmed to express uncertainty when prompts go beyond training knowledge or reasoning ability (41). Even if Strachan et al. (2024) also tested novel English items for their complex ToM tasks and reported no significant differences for the Faux Pas task between new and original items for GPT-4, only worse performance for GPT-3.5; albeit, the form and the language of the items was still similar to the established items, so possibly an entirely new kind of items and another language would be a better comparison. Altogether, it seems that GPT might have marked its higher uncertainty in responses to German prompts because of a basically differing body of training data that could lead to systematical differences.

**Social Stories Questionnaire**

In the Social Stories Questionnaire task GPT-4 (67% German; 63% English) reached a level comparable to neurotypical adults (70% Females; 60% Males), whose results were included from Lawson et al. (2004). In contrast, the performance of GPT-3.5 (28% German; 20% English) was below the level reported for autistic males (50%). These results were independent of the language in which the items were prompted. For this task there was no evaluation of uncertainty markers because of the nested structure of the task, where the response format allowed only dichotomous answers. As – to our knowledge – until now no other study evaluated the Social Stories Questionnaire with a LLM it is not possible to compare these results to other research. However, given that it includes both subtle and blatant social faux pas, the task itself may be comparatively demanding, as even neurotypical adults' accuracy rates are not particularly high (60-70%, Lawson et al., 2004), therefore, GPT-4's results seem adequate. In a more detailed analysis (S3 appendix), GPT-4's performance suggests that it can capture many of the relevant social inferences but still struggles with the more nuanced items, which likely require a deeper integration of contextual and mental state information. In contrast, GPT-3.5's substantially lower accuracy indicates broader limitations in representing and interpreting social meaning that were addressed in the newer versions of GPT.

**Story Comprehension Test**

In this task, GPT-4 again outperformed GPT-3.5 (56% German and English), however, its performance in German language (89%) even exceeded neurotypical adults (64%; 16), while its responses to the English prompts (64%) were on a comparable level to neurotypical adults and above neurotypical adolescents (50%; 16). These results suggest again a high level of proficiency of GPT-4, but there are also no previous studies that tested this task with a LLM known to us. However, as to whether GPT-4 in German responses even outperformed neurotypical adults remains relatively unclear at present. The fact that GPT-4 had more training in the English language (Lommel, 2024) seems to be of no consequence here. These unexplained differences emphasize the need for further cross-linguistic comparisons with LLM applications.

Regarding uncertainty markers, GPT-3.5 used such expressions in approximately half of its responses to the German prompts, whereas GPT-4 used them in about one third of its responses to both the English and German prompts. The very high proportion of uncertainty markers in GPT-3.5's responses to the German prompts might be a consequence of limited training (effects) in German language and general limitations in social understanding as discussed above. However, since GPT-4's responses to the English prompts also contained a considerable number of such markers this phenomenon cannot be attributed solely to language specific training effects (Lommel, 2024). When comparing this relatively high proportion of uncertainty markers to the approximately 6 percent found in responses by relatively large human samples (*n*=60), both neurotypical adults and adolescents, from our previous research (16), the application of GPT's answers in real-world scenarios needs further study. As generally, the evaluation of uncertainty markers was not a topic of ToM research until now, it is unknown whether the rates found by us are at an average level for neurotypical human responses or not. It must be the aim of future studies to determine this and to evaluate, whether usage of uncertainty markers is perceived as misleading in social situations by neurotypical and neurodivergent individuals or whether it simulates human behaviour and is therefore perceived differently.

**Strengths and Limitations**

An important strength of the present study is the systematic evaluation of GPT as a bench-mark LLM across multiple well-established higher-order ToM tasks in both English and German. This cross-linguistic approach increases ecological validity, as it mirrors real-world usage scenarios in which users interact with language models in different native languages. Furthermore, as mentioned before, the tasks and their correct answers could have appeared during model training in the English language, potentially inflating performance estimates (31). Therefore, evaluating the model in German, where exposure during pretraining is likely to be lower, allows for a more robust assessment of its generalizable ToM abilities. In contrast to previous studies that reported prompting-related issues (26), we implemented prompt engineering to reduce hallucinations or inaccurate responses. GPT was instructed to select the single most plausible answer and to respond concisely and precisely. In addition, the application of double-rating procedures with high interrater reliability ensured a reliable assessment of accuracy, above all with a more extensive evaluation of responses that went beyond a coding of correct vs. incorrect. This seems to add to ecological validity in our assessment. A further methodological contribution is the systematic coding of uncertainty markers, providing novel insights into the models' communicative style and its potential implications for clinical applications.

Despite these strengths, several limitations must be acknowledged. First, the analysis was confined to GPT-3.5 and GPT-4. Although these models are among the most widely used, the extent to which the results transfer to other LLM's (e.g., Gemini, Claude, Perplexity) remains an open question. In Strachan et al.'s (2024) study LLaMA performed better than GPT in the Faux Pas test, but not in the other complex ToM tasks that were tested.Second, despite careful efforts to mitigate training contamination through translation into German, the possibility of overlap between test material and pretraining corpora, particularly in the English version, cannot be ruled out, as discussed above. Fourth, the evaluation was restricted to text-based interactions. However, ToM in real-life communication relies not only on linguistic information but also on a broad range of non-linguistic cues such as prosody, facial expressions, and gestures (42). Because these modalities were not captured in the present design, the ecological validity of the findings is limited. Finally, the interpretation of uncertainty markers is complex. Their frequent occurrence may reflect genuine uncertainty and thus transparency of the models, but could equally be attributable to stylistic hedging strategies acquired during training (Lommel, 2024). This ambiguity complicates conclusions about the functional role of such markers and their potential utility in applied settings.

It should also be noted that the present study focused exclusively on GPT-3.5 and GPT-4. Since the completion of data collection, newer versions of large language models such as GPT-5 have been released. Although the extent of their improvements in Theory of Mind tasks remains to be systematically evaluated, the rapid pace of model development highlights the need for continuous reassessment of ToM capabilities in successive model generations.

**Conclusion**

Altogether we found comparably high accuracy rates of GPT-4 across all tasks and languages, which exceeded the results of previous studies. This was particularly evident in the Faux Pas Test (28). GPT-4's results were on par with those of neurotypical adults and exceeded those of neurodivergent adults, where previous studies had reported results in two out of three tasks. However, GPT-3.5 showed continuously worse performance than GPT-4, achieving scores below those of neurotypical adults. This reflects the limitations addressed in subsequent versions of GPT. As different reasons could account for the contrasting results by Strachan et al. (2024), future studies should not only use a variety of ToM tasks, but also different languages to control for training effects of the LLM.

Regarding the evaluation of uncertainty markers, no previous research had evaluated LLM's use of those, so our results could only be compared to research based on human subjects. The usage of uncertainty markers by GPT was considerably higher than that of humans; therefore, the high proportion of uncertain statements in GPT's responses emphasises the need for further studies that evaluate practical applications of LLM's. Consequently, it is unclear, if GPT would show worse results in real-world scenarios, so this must be the subject of future studies.

Overall, the results of GPT's ToM abilities in complex social vignettes are promising; however, practical applications require further investigation - particularly with respect to whether the model can provide the adequate and context-specific communication needed by neurodivergent individuals in ambiguous social situations (8). At the same time, these findings prompt philosophical reflection on the nature of GPT's seeming social understanding (21). However, the focus of this current research was not to determine whether a LLM, here GPT, possesses genuine ToM, merely to evaluate the quality of its simulation of this applied social reasoning and whether it could provide helpful assistance in real-world scenarios. More study is needed particularly regarding the usage of uncertain statements and whether neurodivergent as well as neurotypical humans find the provided assistance as helpful or not. This highlights the difficulty of determining what it truly means to "understand" another mind.

THEORY OF MIND AND CHATGPT

**Appendix**

**S1 Appendix. Python Code.**

```python
from datetime import datetime

from threading import Thread

from time import sleep

from typing import Dict, List

import gspread

import openai

from google import auth

openai.api_key = "<PUT_YOUR_OPENAI_API_KEY_HERE>"

MODELS = ["gpt-4", "gpt-3.5-turbo"]   # these are just defaults, you can
choose any models available

# spreadsheet must contain a questions worksheet with each question in a
separate cell in the first column
# and for each model in the list of models above, a worksheet with the name
of the model
SPREADSHEET_ID = "<PUT_YOUR_GOOGLE_SPREADSHEET_ID_HERE>"

QUESTIONS_WORKSHEET = "questions"

QUERIES_PER_QUESTION = 10

REQUEST_TIMEOUT_SECONDS = 60

TARGET_SCOPES = [

  'https://www.googleapis.com/auth/accounts.reauth',

  'https://www.googleapis.com/auth/drive',

  'https://www.googleapis.com/auth/drive.file',
```

```python
    'https://www.googleapis.com/auth/drive.readonly',

    'https://www.googleapis.com/auth/spreadsheets',

    'https://www.googleapis.com/auth/spreadsheets.readonly'

]

def read_questions(sheet: gspread.Spreadsheet) -> List[str]:

    worksheet = sheet.worksheet(QUESTIONS_WORKSHEET)

    return worksheet.col_values(1)

def query(question: str, model: str) -> str:

    return openai.ChatCompletion.create(model=model,

                    messages=[{"role": "user", "content": question}],

request_timeout=REQUEST_TIMEOUT_SECONDS)["choices"][0]["message"]["content"
]

def get_answers(questions_list: List[str], model_to_answers: Dict[str,
List[List[str]]], model: str) -> None:

    for i in range(QUERIES_PER_QUESTION):

        for j in range(len(questions_list)):

            retries = 0

            while True:

                try:

                    answer = query(questions_list[j], model)

                    break

                except Exception as e:

                    retries += 1

                    print(f"error while querying question {j} iteration {i} model
{model}, retrying for {retries} time")

                    print(f"error: {e}")

                    sleep(1)
```

```python
        model_to_answers[model][i].append(answer)

        print(f"finished question {j} for iteration {i} model {model}")

        print(answer)

def write_answers(model_to_answers: Dict[str, List[List[str]]], sheet:
gspread.Spreadsheet) -> None:
  for model, answers_list in model_to_answers.items():
    worksheet = sheet.worksheet(model)
    for i in range(len(answers_list)):
      worksheet.append_row(values=answers_list[i], table_range=f"A{i +
1}:AW{i + 1}")

start = datetime.now()
print("starting to read questions...")
default_creds, _ = auth.default(scopes=TARGET_SCOPES)
gc = gspread.authorize(default_creds)
sh = gc.open_by_key(SPREADSHEET_ID)
questions = read_questions(sh)
print(f"read {len(questions)} questions!")
answers = {model: [[] for _ in range(QUERIES_PER_QUESTION)] for model in
MODELS}
threads = []

for model in MODELS:
  print(f"starting thread for model {model}")
  threads.append(Thread(target=get_answers, daemon=True, args=(questions,
answers, model)))
  threads[-1].start()

print(f"waiting for {len(threads)} threads to finish...")
count = 0
```

THEORY OF MIND AND CHATGPT

```python
for t in threads:
    try:
        t.join()
    except KeyboardInterrupt:
        pass
    count += 1
    print(f"{count} threads finished")

print("writing down answers...")
write_answers(answers, sh)
end = datetime.now()
print(f"finished in {end - start}")a
```

**S2 Appendix. Prompts ChatGPT.**

**Prompt example for the Story Comprehension Test in German and English:**

**German:** Bitte lies die Geschichte aufmerksam durch und beantworte die Frage so ausführlich und genau wie möglich. Bitte beschränke dich auf eine, die wahrscheinlichste Antwortmöglichkeit. Monikas Tochter Lara spielt gern nach der Schule mit ihren Freunden, aber ihr wurde gesagt, dass sie um 5 Uhr zu Hause sein soll. Manchmal kommt Lara zu spät und Monika macht sich Sorgen, dass etwas passiert sein könnte. An diesem Abend sagt Monika zu Lara „Wenn du nicht um 5 zu Hause bist, bekommst du kein Abendbrot." Warum hat Monika das gesagt?

**English:** Read the story and answer the question as precisely and detailed as possible. Please give a single most likely answer: Monica's daughter Laura likes to play with her friends after school, but she was told to be home by 5 o'clock. Sometimes Laura is late and Monica worries that something might have happened. That evening Monika says to Lara "If you're not home by 5, you won't get dinner." Why did Monica say that?

THEORY OF MIND AND CHATGPT

<div align="center">

**S3 Appendix. Results Subtle and Blatant Faux Pas.**

</div>

**Correct Identification of Subtle Faux Pas**

The detection of subtle items in the Social Stories Questionnaire was particularly infrequent. In version 3.5, the detection of subtle items was notably low, with a mean score of $M = 1.50$ ($SD = 0.55$) for German prompts and $M = 1.0$ ($SD$ could not be computed, as only one observation was available) for English prompts. In version 4, GPT's performance was better. English prompting had a mean score of $M = 3.50$ ($SD = 0.85$). German prompting achieved a mean score of $M = 3.60$ ($SD = 0.84$). These scores indicate a notable improvement over version 3.5.

In comparison to the overall scores and the values for the blatant items shown subsequently, the subtle items appear to be identified correctly less frequently.

**Results for the Correct Identification of blatant Faux Pas**

A similar pattern of language and version effects for the identification of subtle faux pas also emerged for blatant faux pas, although overall accuracy was higher in version 3.5, (German: $M = 4.8$, $SD = 0.789$; English: $M = 4.0$ $SD = 0.816$) in comparison to the performance of GPT-4 (German: $M = 9.90$, $SD = 0.32$, English: $M = 9.10$, $SD = 0.88$).